\documentclass[twocolumn,showpacs,preprintnumbers,amsmath,amssymb]{revtex4}

\usepackage{graphicx}
\usepackage{dcolumn}
\usepackage{bm}

\begin{document}

\title{The orientation and magnitude of the orbital precession
velocity of a binary pulsar system with
double spins}
\author{B.P. Gong}

\affiliation{ Department of Astronomy, Nanjing University, Nanjing
210093,PR.China}

\email{bpgong@nju.edu.cn}


\begin{abstract}
The measurability of the spin--orbit (S--L) coupling induced
orbital effect is dependent on the orientation and magnitude of
the orbital precession velocity, ${\bf \Omega}_0$. This paper
derives ${\bf \Omega}_0$ in the case that both spins in the binary
system contribute to the spin--orbit (S--L) coupling, which is
suitable for the most popular binary pulsars, Neutron star--White
Dwarf star (NS--WD) binaries (as well as for NS--NS binaries).
This paper shows that from two constraints, the conservation of
the total angular momentum and the triangle formed by the orbital
angular momentum, ${\bf L}$, the sum the spin angular momenta of
the two stars, ${\bf S}$, and the total angular momentum, ${\bf
J}$, the orbital precession velocity, ${\bf \Omega}_0$,  along
${\bf J}$ is inevitable. Moreover, by the relation, $S/L\ll 1$,
which is satisfied for a general binary pulsar, a significant
${\bf \Omega}_0$ (in magnitude) is inevitable, 1.5 Post Newtonian
order (PN). Which are similar to the case of one spin as discussed
by many authors.
However unlike the one spin case, the magnitude of
the precession velocity of ${\bf \Omega}_0$ varies significantly
due to the variation of the sum the spin angular momenta of the
two stars, ${\bf S}$, which can lead to significant secular
variabilities in binary pulsars.
\end{abstract}
 \draft 

\maketitle

\section{Introduction}
Barker $\&$ O'Connell\cite{bo} derived the first gravitational
two-body equation with spins.  In which the precession of the
angular momentum vector, ${\bf L}$, responsible for the S--L
coupling is expressed as around a combined vector of the spin
angular momenta of the pulsar, ${\bf S}_1$, and its companion
star, ${\bf S}_2$ (${\bf S}={\bf S}_1+{\bf S}_2$). Which is
insignificant, 2PN, typically $10^{-4}$deg per year, and therefore
ignorable in pulsar timing measurement. But notice that this
ignorable orbital precession velocity, ${\bf \Omega}_0$, is
expressed as along a combined vector by ${\bf S}_1$ and ${\bf
S}_2$, which is not static to the line of sight (LOS). Therefore,
the measurability of ${\bf \Omega}_0$ need to be further
investigated.

Since the precession of the pulsar spins (${\bf S}_1$ and ${\bf
S}_2$) around ${\bf L}$ (and ${\bf J}$) is significant, 1.5 PN
(typically $1$deg per year) by Barker $\&$ O'Connell two--body
equation, and the precession of ${\bf S}_1$ relative to the line
of sight (LOS) has been measured through the structure parameters
of pulsar profile\cite{kr,wt},
Therefore the small precession velocity of ${\bf L}$ with respect
to a combined vector by ${\bf S}_1$ and ${\bf S}_2$ doesn't mean
that it is also small relative to LOS.

 For binary pulsars the change of the orbital period
due to the gravitational radiation is 2.5PN; whereas  the geodetic
precession of the pulsar is 1.5PN. Therefore, the influence of
gravitational radiation on the motion of a binary system can be
ignored. And the total angular momentum, ${\bf J}$ (${\bf J}={\bf
L}+{\bf S}_{1}+{\bf S}_{2}$), can be treated as invariable both in
magnitude and direction. Therefore after counting out the proper
motion effect, ${\bf J}$ is at rest to LOS. So to make sense in
observation, the precession velocity, i.e., ${\bf \Omega}_0$,
should be expressed as around ${\bf J}$.

In the study of observational effect of the orbital precession due
to the coupling of the spin induced quadruple moment of the
companion star with the orbital angular momentum\cite{sb,lai,wex},
 the precession of ${\bf L}$ is expressed as
relative the total angular momentum vector, ${\bf J}$. And the
orbital velocity obtained can explain some secular variabilities
measured. However this model might not suitable for neutron
star-white dwarf (NS-WD) and neutron star-neutron star (NS-NS)
binaries which have much smaller quadruple moment.

On the other hand, Apostolatos et al\cite{apo}, Kidder\cite{kid}
as well as Wex and Kopeikin\cite{wex} studied the precession of
${\bf L}$ as relative to ${\bf J}$ in discussing the spin-orbit
(S--L) coupling in the NS-BH binary systems. Similar to the former
quadruple-orbit coupling, this S--L coupling only considers the
spin (quadruple) of the companion star (BH), the spin of the
pulsar is ignored. The difference is that the former Q--L coupling
corresponds to 2\,PN along and ${\bf J}$, whereas the latter S-L
coupling corresponds to 1.5 PN along ${\bf J}$. Therefore, the
latter is easier to explain the measurements of NS-WD and NS-NS
binaries, which have much smaller quadruple moments.

The S--L coupling in the one spin case predicts a static
precession of ${\bf L}$, or ${\bf \Omega}_0$ is unchanged both in
magnitude and direction. Therefore, the orbital effect induced in
such case can be completely absorbed the orbital parameters such
as orbital period, $P_b$, and the precession of periastron,
$\dot{\omega}$. However, this model cannot explain the significant
secular variabilities measured in many binary pulsars. Therefore,
it actually has not been used in pulsar timing measurement.

This paper studies the precession velocity of  ${\bf L}$ around
the static direction, ${\bf J}$, in the a general case, two spins.
It comes to the following conclusion: (a) as that of one spin
case, the three vectors, ${\bf L}$, ${\bf J}$ and ${\bf S}$ are
still in the same plane at any instant; (b) as that of one spin
case, the precession velocity of ${\bf L}$ around ${\bf J}$ is
significant (measurable for many binary pulsars);
(c) unlike the one spin case, the magnitude of the precession
velocity of ${\bf \Omega}_0$ varies significantly due to the
variation of the sum the spin angular momenta of the two stars,
${\bf S}$, which can naturally explain the significant secular
variabilities measured in binary pulsars\cite{go}.

This paper is arranged as follows. Section II introduces the
orbital velocity of Barker and O'Connell\cite{bo} and that of one
spin equation by Kidder\cite{kid}, Wex and Kopeikin\cite{wex}. The
measurability of them are also discussed and compared. Section III
derived the orbital velocity as around ${\bf J}$ in the  general
(two spins) case. Section IV gives the derivatives of ${\bf
\Omega}_0$, responsible for the significant secular variabilities.

\section{Orbital precession velocities by Barker $\&$
O'Connell and other authors} According to the gravitational
two-body equations\cite{bo},
 the motion of the a binary system can be described by three vectors, the spin angular momentum of the
 pulsar, ${\bf S}_{1}$, its companion star, ${\bf S}_{2}$, and the orbital angular momentum, ${\bf L}$.
  The secular result for  ${\bf S}_{1}$  is
$$ {\bf \Omega}_{1}= \frac {GL
(4+3m_{2}/m_{1})}{2c^{2}a_{R}^{3}(1-e^{2})^{3/2}}\,{\bf n}_{0}$$
\begin{equation}
\label{e1x1} +\frac {GS_{2}}{2c^{2}a_{R}^{3}(1-e^{2})^{3/2}}
\left[{\bf n}_{2}-3({\bf n}_{0}\cdot{\bf n}_{2})\,{\bf
n}_{0}\right] \ ,
\end{equation}
where ${\bf n}_{0}$, ${\bf n}_{1} $  and ${\bf n}_{2}$ are unit
vectors of the orbital angular momentum, the spin angular momentum
of the pulsar and its companion star, respectively,
  $ m_{1}$, $m_{2}$ are the masses of pulsar and
 the companion star, respectively, $e$ is the eccentricity of the orbit,
 $a_{R}$ is the semi-major axis.
The first term corresponds to the geodetic (de Sitter) precession
which represents the precession of ${\bf S}_{1}$ around ${\bf L}$,
since  $L\propto a_{R}^{1/2}$, the first term actually corresponds
to $a_{R}^{-5/2}$ (1.5 PN); and the second term represents the
Lense-Thirring precession, ${\bf S}_{1}$ around ${\bf S}_{2}$,
which corresponds to $a_{R}^{-3}$ (2 PN). ${\bf \Omega}_{2}$ can
be obtained by exchanging the subscript 1 and 2 at the right side
of Eq($\ref{e1x1}$).

The precession of the orbital angular momentum, ${\bf
\Omega}_{0}$, can be given as follows\cite{bo}:
\begin{equation}
\label{we1} {\bf \Omega}_{0}={\bf \Omega}_{PN} +{\bf
\Omega}_{L}+{\bf \Omega}_{Q}\,,
\end{equation}
where ${\bf \Omega}_{PN}$ is the relativistic periastron advance,
${\bf \Omega}_{L}$ is the geodetic precession caused by S-L
coupling, and ${\bf \Omega}_{Q}$ is the precession due to the
Newtonian coupling of the orbital angular momentum vector to the
quadruple moment of the two bodies. For a general binary pulsar
system, like NS--NS and NS--WD  binaries, ${\bf \Omega}_{Q}$ is
ignorable, 2.5PN, ${\bf \Omega}_{PN}$ is significant, 1.5PN,  but
it has no contribution to the S--L coupling, due to it is along
 ${\bf L}$. Therefore, only ${\bf \Omega}_{L}$ (2
PN) contributes to the S--L coupling. By Barker and
O'Connell\cite{bo},
\begin{equation}
\label{e1x2} {\bf \Omega}_{L}
=\sum\limits_{\alpha=1}^{2}\frac{GS_{\alpha}(4+3m_{\alpha+1}/m_{\alpha})}
{2c^{2}a_{R}^{3}(1-e^{2})^{3/2}}\, \left[{\bf n}_{\alpha}-3({\bf
n}_{0}\cdot{\bf n}_{\alpha}){\bf n}_{0}\right] \ ,
\end{equation}
where $\alpha+1$ meant modulo 2 ($2+1=1$), Eq($\ref{e1x2}$) means
that ${\bf L}$ precesses around a vector combined by ${\bf S}_{1}$
and ${\bf S}_{2}$ with a very small velocity (2PN), and the
precession of ${\bf L}$ due to geodetic precession is thus
considered negligible to pulsar timing measurement.

However since ${\bf S}_{1}$ and ${\bf S}_{2}$ and thus the
combined vector by ${\bf S}_{1}$ and ${\bf S}_{2}$  precess
rapidly (1.5PN) with respect to LOS, therefore from the small
precession velocity ${\bf L}$ to the combined vector of ${\bf
S}_{1}$ and ${\bf S}_{2}$ (2PN) given by Eq($\ref{e1x2}$), we
still cannot conclude that ${\bf L}$  also precesses with respect
to LOS slowly.

Actually in the study of the observational effect of the Q--L
coupling, ${\bf \Omega}_{Q}$ is transformed to the direction of
${\bf J}$.  And the value of  ${\bf \Omega}_{Q}$ along ${\bf J}$
is about L/S times larger than that of the previous one, thus
${\bf \Omega}_{Q}$ along ${\bf J}$ corresponds to 2PN instead of
the previous 2.5PN.

Similarly if orbital velocity corresponding to the S--L coupling
 is also expressed as along ${\bf J}$,
then it would also be L/S times the velocity ${\bf \Omega}_{L}$
given by Eq($\ref{e1x2}$), which would be 1.5PN instead of 2PN.
Then it should be comparable to ${\bf \Omega}_{PN}$ in magnitude
and measurable.

Actually this work has been do by many authors\cite{apo,kid,wex},
in the special case of one spin. From the conversation of the
total angular momentum, we have $\dot{{\bf J}} = 0$, which can be
written,
\begin{equation}
\label{e1} {\bf \Omega_{0}}\times {\bf L}=-{\bf \Omega_{1}}\times
{\bf S_{1}}-{\bf \Omega_{2}}\times{\bf S_{2}}
 \ \,.
\end{equation}
Notice that as defined by Barker and O'Connell\cite{bo} and
Apostolatos et al\cite{apo}, ${\bf L}=\mu M^{1/2}r^{1/2}{\bf
n}_0$, where $M=m_1+m_2$, and $\mu=m_1m_2/M$.

In the special case that one spin angular momentum is ignorable,
i.e., ${\bf S_{2}}=0$, as discussed by Kidder\cite{kid}, Wex and
Kopeikin\cite{wex}, (Apostolatos et al\cite{apo} discussed the
special case that ${\bf S_{1}}\cdot{\bf S_{2}}=0$),  $S$ is a
constant in magnitude. And thus ${\bf L}$ and ${\bf S}$ precess
about the fixed vector ${\bf J}$ at the same rate with a
precession frequency approximately\cite{kid}
\begin{equation}
\label{e1aa4}\Omega_{0}=\frac {|{\bf J}|}{2r^{3}}
(4+\frac{3m_{2}}{m_{1}}) \ .
\end{equation}
Note that Eq($\ref{e1aa4}$) uses $G=c=1$ and
$r=a_{R}(1-e^{2})^{1/2}$.
 Since ${\bf L}$, ${\bf J}$ and ${\bf S}$ forms a triangle,
therefore, the three vectors are in the same plane. Or more
clearly, ${\bf L}$ and ${\bf S}$ are at the opposite sides of
${\bf J}$ at any instant (${\bf J}$ is invariable).

Notice the velocity given by Eq($\ref{e1aa4}$) is relative to the
static direction ${\bf J}$ (which makes sense in observation), and
the magnitude of Eq($\ref{e1aa4}$) is 1.5PN, which means the
precession velocity of ${\bf L}$ around ${\bf J}$ is much larger
than that of  ${\bf L}$ around a combined vector by ${\bf S}_1$
and ${\bf S}_2$ given by Eq($\ref{e1x2}$).

If Eq($\ref{e1aa4}$) is applied to pulsar timing measurement, the
static precession of ${\bf L}$ around ${\bf J}$ can be completely
absorbed by parameters like the orbital period, $P_b$, and the
advance of precession of periastron, $\dot{\omega}$. Therefore,
the effects of Eq($\ref{e1aa4}$) might be unmeasurable, although
it is significant.

However, if both spins are considered, then the precession of
${\bf S}_1$ and ${\bf S}_2$ will lead to a variable ${\bf S}$
(both in direction and magnitude), and therefore, through S--L
coupling, the orbital velocity, ${\bf \Omega}_{0}$ varies in
magnitude. Which cannot be completely  absorbed by parameters,
like, $P_b$ and $\dot{\omega}$. And it is this variation that
automatically explains the secular variabilities measured in many
binary pulsars\cite{go}.

Therefore, it is necessary to derive an orbital precession
velocity that is relative to ${\bf J}$ and including two spins.

\section{orbital precession velocity in general case}
\subsection{The scenario of the motion of ${\bf L}$, ${\bf J}$ and ${\bf S}$}
This section derives orbital precession velocity as around ${\bf J
}$ in the case that ${\bf S_{1}}\neq 0$ and ${\bf S_{2}}\neq 0$,
so that it can be applied to general NS--WD and NS--NS binaries.
 Eq($\ref{e1}$) can be
rewritten as
\begin{equation}
\label{e1ax} {\bf \Omega_{0}}\times {\bf L}=-{\bf
\Omega_{2}}\times {\bf S}-({\bf \Omega_{1}}-{\bf
\Omega_{2}})\times{\bf S_{1}} \ \,.
\end{equation}
Notice that when ignoring the terms over 2PN, ${\bf \Omega}_{1}$
and ${\bf \Omega}_{2}$ are along ${\bf n}_{0}$. The first term at
right side of  Eq($\ref{e1ax}$) represents a torque that is
perpendicular to the plane determined by ${\bf L}$ and ${\bf S}$.
And the second term  at right side of Eq($\ref{e1ax}$) also has a
component in the same direction as the first one, which can be
written as, $(\Omega_{1}-\Omega_{2})S^{\parallel}_{1}\sin
\lambda^{\parallel}_{LS_{1}}$, in magnitude. Where
$S^{\parallel}_{1}$ corresponds the component of  ${\bf S}_{1}$ in
the plane determined by  ${\bf S}$ and  ${\bf J}$, which satisfies
$S^{\parallel}_{1}=S_{1}\cos \eta_{SS_{1}}$, and
$\lambda^{\parallel}_{LS_{1}}$ is the misalignment angle between
$S^{\parallel}_{1}$ and ${\bf J}$.
$\eta_{SS_{1}}=\eta_{S_{1}}-\eta_{S}$ ($\eta_{S_{1}}$ and
$\eta_{S}$ are precession phases of ${\bf S}_{1}$ and ${\bf S}$
respectively).

Obviously ${\bf S}$ should precess at the velocity which is equal
to that ${\bf L}$ precess around ${\bf J}$, whereas,
Eq($\ref{e1ax}$) uses ${\bf \Omega_{2}}$ to replace this velocity,
therefore, an extra term
($(\Omega_{1}-\Omega_{2})S^{\parallel}_{1}\sin
\lambda^{\parallel}_{LS_{1}}$) is needed to make up the
difference. The torques made by these two terms can be given,
\begin{equation}
\label{e1ax1} (\tau_{1})_{R}=[\Omega_{2}\sin\lambda_{LS}+
(\Omega_{1}-\Omega_{2})\frac{S^{\parallel}_{1}}{S}\sin
\lambda_{LS_{1}}^{\parallel}]S
 \ ,.
\end{equation}
$(\tau_{1})_{R}$ is equivalent to a combined one,
$\Omega_{S}{S}\sin\lambda_{LS}$. The second term at right hand
side of  Eq($\ref{e1ax}$) can produce a torque that is
perpendicular to that of the first term. Which can be written,
\begin{equation}
\label{e1ax2} (\tau_{2})_{R}=
(\Omega_{1}-\Omega_{2})S^{\perp}_{1}\sin \lambda^{\perp}_{LS_{1}}
 \ ,
\end{equation}
where $S^{\perp}_{1}$
is the component of the ${\bf S}_{1}$, which is at the plane
 perpendicular to the plane determined by
${\bf L}$ and ${\bf S}$ instantaneously. Which produces a torque,
$(\tau_{2})_{R}$, that is perpendicular to $(\tau_{1})_{R}$,
obviously in the one spin case, $(\tau_{2})_{R}=0$. Actually
Eq($\ref{e1ax2}$) can also be regarded as ${\bf S}$ precesses
around a vector that is perpendicular to the plane determined by
${\bf L}$ and ${\bf S}$ with velocity $(\tau_{2})_{R}/S$
instantaneously.

$(\tau_{2})_{R}\neq 0$ (in the general case) indicates that the
left hand side of  Eq($\ref{e1}$) and Eq($\ref{e1ax}$) must make a
corresponding torque to balance with $(\tau_{2})_{R}$, or the
orbital precession velocity, ${\bf \Omega}_0$, much have a
component that is perpendicular to the plane determined by ${\bf L
}$ and ${\bf S}$  to balance with $(\tau_{2})_{R}$. Then it seems
that  the three vectors may not be in the same plane (or ${\bf J
}$ is not in the  plane determined by ${\bf L }$ and ${\bf S}$) ,
due to  $(\tau_{2})_{R}\neq 0$.

However this situation will never happen because the constraint,
${\bf J}={\bf L }+{\bf S}$, must be satisfied at any instant in
the two spins case also. Which demands that ${\bf J}$, ${\bf L }$
and ${\bf S}$ must be in the same plane at any instant.
Simultaneously, the constraint, $\dot{{\bf J}}=0$, demands that
${\bf J}$ is unchanged both in magnitude and direction. Therefore,
${\bf L }$ and ${\bf S}$ have to be at the opposite side of ${\bf
J}$ at any instant to satisfy to the two constraints at the same
time. In other words, the plane determined by ${\bf L }$ and ${\bf
S}$ rotates around a fixed vector, ${\bf J}$.

Then what is the role of the torque, $(\tau_{2})_{R}$, in the
motion of the three vectors? It can cause a precession of ${\bf
S}$ around a vector perpendicular to the plane determined by ${\bf
L }$, ${\bf J}$ and ${\bf S}$ (plane LJS) instantaneously, which
can change the angle between ${\bf L }$ and ${\bf S}$. Similarly
${\bf L }$ also precesses around a direction perpendicular to the
plane LJS instantaneously, which changes the angle between ${\bf L
}$ and ${\bf J}$.

Thus $(\tau_{2})_{R}$ can only changes the misalignment angles of
${\bf L}$ and ${\bf S}$ with respect to ${\bf J}$, it cannot
influence the scenario that ${\bf L }$ and ${\bf S}$ precess
rapidly at opposite side of ${\bf J}$ at any moment. In the one
spin case $(\tau_{2})_{R}=0$, the misalignment angles between
${\bf S }$ and ${\bf L}$ as well as ${\bf L }$ and ${\bf J}$ are
unchanged.

Therefore the difference between one spin and two spins is that
the former corresponds to a simple precession of ${\bf L }$ and
${\bf S }$ around ${\bf J}$, but the latter corresponds to a
nutation superimposed on an overall precession. While the common
point is that at any instant, the three vectors are in the same
plane, and moreover, ${\bf L }$ and ${\bf S}$ are at the opposite
side of ${\bf J }$.

As discussed above from the two constraints, ${\bf J}={\bf L
}+{\bf S}$ and $\dot{{\bf J}}=0$, we can conclude that the plane
determined by ${\bf L }$ and ${\bf S}$ precesses around a fixed
vector, ${\bf J}$, rapidly (1.5PN). In other words, the precession
velocity of ${\bf L }$ and ${\bf S}$ (${\bf \Omega}_0$ and ${\bf
\Omega}_S$) is along ${\bf J}$. One can also choose vectors other
than ${\bf J}$ as axis that  ${\bf L }$ and ${\bf S}$ precess
around. However that would be like using a planet instead of the
Sun as the center of a reference frame to explain the motion of
the solar system.

Actually the scenario of the motion of vectors in a binary system
have been discussed by authors, Smarr and Blandford\cite{sb},
Hamilton and Sarazin\cite{hs}. From which one can easily deduce
that the small velocity of  ${\bf L }$ around a combined vector by
${\bf S}_1$ and ${\bf S}_2$ doesn't influence that the three
vectors, ${\bf L }$, ${\bf S}_1$ and ${\bf S}_2$ precess rapidly
around ${\bf J}$. In other words, the precession velocity of ${\bf
L }$ around ${\bf J}$ can be significant and therefore measurable.

Since Barker and O'Connell's two-body equation satisfies the two
constraints, ${\bf J}={\bf L }+{\bf S}$ and $\dot{{\bf J}}=0$.
Therefore, Barker and O'Connell's two-body equation actually means
that the orbital precession velocity, ${\bf \Omega}_0$, is along
${\bf J}$. However the misalignment angle between ${\bf L}$ and
${\bf J}$ is arbitrary in Barker and O'Connell's two-body
equation, therefore, it is impossible to obtain ${\bf \Omega}_0$
along ${\bf J}$ without further constraint.

\subsection{The magnitude of ${\bf \Omega}_0$}
Fortunately there is a  constraint, $S/L \ll 1$, that must be
satisfied for a general binary pulsar. $S/L \ll 1$ can impose
strong constraint on $\lambda_{LJ}$ (misalignment angle between
${\bf J}$ and ${\bf S}$), and therefore, the magnitude (also
direction) of ${\bf \Omega}_0$.

By the triangle ${\bf J}={\bf L }+{\bf S}$ and the relation $S/L
\ll 1$, one can know that the misalignment angle between  ${\bf
L}$ and ${\bf J}$ must be very small, $\lambda_{LJ}\ll 1$. Thus no
matter ${\bf L}$ participates in what ever motions, precession and
nutation, ${\bf L}$ must always be very close to ${\bf J}$.

As discussed above, $(\tau_{1})_{R}$ and $(\tau_{2})_{R}$
correspond to torques caused by components of ${\bf S}_{1}$ and
${\bf S}_{2}$ that is in the plane LJS and perpendicular to it
respectively. Then  ${\bf L}$ must make a torque that equals
$(\tau_{1})_{R}$ in magnitude. Then ${\bf L}$ can react to
$(\tau_{1})_{R}$ in two ways:

(a) ${\bf L}$ can precession around a vector combined by ${\bf
S}_{1}$ and ${\bf S}_{2}$ (which is the previous treatment of
Eq($\ref{e1x2}$));

(b) ${\bf L}$ precesses around ${\bf J}$, with the opening angle
of the precession cone $\lambda_{LJ}\approx S/(\rho L)$
($\rho=1/\sin\lambda_{JS}$).

Mathematically,  both (a) and (b) are allowed.  (a) doesn't
contradictory with (b), since that when ${\bf L}$ has certain
angle with respect to the vector combined by ${\bf S}_{1}$ and
${\bf S}_{2}$ (usually this angle is large), the angle between
${\bf L}$ and ${\bf J}$ can be arbitrary. Because the magnitude of
${\bf L}$ and ${\bf S}$ can be both comparable and very different
($S/L\ll 1$) in Barker and O'Connell's equation.

In (b), although the angle between ${\bf L}$ and ${\bf J}$ is very
small, the angle between  ${\bf L}$ and the vector combined by
${\bf S}_{1}$ and ${\bf S}_{2}$ (or ${\bf S}$) can still be
arbitrary. Therefore, (b) doesn't contradictory with (a). The
difference is that (b) uses one more constraint (which must be
satisfied by a general binary pulsar) than (a), that is
$\lambda_{LJ}\ll 1$. Thus for binary pulsars, the orbital velocity
obtained from (b) is more closer to the true one than that of (a).
Moreover, the result from (a) has no observational means (not at
rest to LOS), whereas the result from (b) make sense in
observation (at rest to LOS).

Therefore, for a general binary pulsar system, especially NS-WD
and NS-NS binaries, only (b) is considered. The left hand side of
Eq($\ref{e1}$) and Eq($\ref{e1ax}$) can be written as
\begin{equation}
\label{e1ax3}
 (\tau_{1})_{L}=|\Omega_{0}{\bf n_{J}}\times L{\bf
n_{0}}|=\Omega_{0}L\sin \lambda_{LJ} \ .
\end{equation}
in which $\Omega_{0}$ denotes the velocity of ${\bf L}$ around
${\bf J}$. By Eq($\ref{e1ax1}$) and Eq($\ref{e1ax3}$), we have the
 precession rate,
\begin{equation}
\label{e1a} \Omega_{0}=\frac{(\tau_1)_R}{L\sin
\lambda_{LJ}}=\rho\Omega_{2}\sin\lambda_{LS}+
\rho(\Omega_{1}-\Omega_{2})\frac{S^{\parallel}_{1}}{S}\sin
\lambda_{LS_{1}}^{\parallel}
 \ \,.
\end{equation}
Note $\rho L\sin \lambda_{LJ}\approx S$. In reaction to the
torque, $(\tau_{2})_{R}$, of Eq($\ref{e1ax2}$), ${\bf L}$ should
precess around a vector, ${\bf n}^{\perp}_{J}$ ( perpendicular to
the plane LJS instantaneously), which corresponds to an opening
angle of the precession cone $\pi/2$ at any moment.
\begin{equation}
\label{e1ax4} (\tau_{2})_{L}=|\Omega^{nu}_{0}{\bf
n}^{\perp}_{J}\times L{\bf
n_{0}}|=\Omega^{nu}_{0}L\sin\frac{\pi}{2} \ ,
\end{equation}
This precession is usually called nutation, superimposed on an
overall precession of the system about the axis of total angular
momentum.  By Eq($\ref{e1ax2}$) and Eq($\ref{e1ax4}$), we have
\begin{equation}
\label{e1a1}
\Omega^{nu}_{0}=\frac{(\tau_2)_R}{L}=\frac{(\Omega_{1}-\Omega_{2})S^{\perp}_{1}\sin
\lambda_{LS_1}^{\perp}}{L}
 \ \,,
\end{equation}
Eq($\ref{e1a}$) corresponds to a rapid precession of ${\bf L}$
around ${\bf J}$ (1.5PN); whereas Eq($\ref{e1a1}$) corresponds to
a slow precession (nutation) of ${\bf L}$ around a vector that is
perpendicular to the plane LJS at any instant (2PN). In other
words, the velocity of Eq($\ref{e1a1}$) derived from the torque,
$(\tau_2)_R$, can only change the misalignment angles,
$\lambda_{LJ}$ and $\lambda_{LS}$, it cannot influence the
direction of ${\bf \Omega}_0$.

In summary, from the constraints, ${\bf L}+{\bf S}={\bf J}$, and
$\dot{\bf J}=0$, the precession velocity ${\bf \Omega}_{0}$ along
${\bf J}$ is inevitable. Further by the constraint, $S/L\ll 1$,
which is correct for a general binary pulsar, a significant
$\Omega_{0}$ (comparable to $\dot{\omega}$ the precession of
periastron) is inevitable.

Therefore, the one spin case and the general case are  similar in
serval aspects.  The orbital precession velocities of a binary
pulsar system are both significant, 1.5PN and both around ${\bf
J}$. Moreover, the three vectors, ${\bf L}$, ${\bf J}$ and ${\bf
S}$ are in the same plane at any instance in both cases. The one
spin case corresponds to a zero nutation velocity, while the two
spins case corresponds to an ignorable nutation velocity of the
orbit plane, 2PN.

However, there is significant difference between one spin and two
spins cases, the former corresponds to a constant S, which means a
static precession of ${\bf S}$ and ${\bf L}$ around ${\bf J}$,
$\Omega_0=const$; whereas, the latter corresponds to a variable S,
and therefore, a variable $\Omega_0$, which is responsible  for
the significant secular variabilities measured in many binary
pulsars. On the other hand, the significant secular variabilities
measured in many binary pulsar seems to support the two spins
case\cite{go}.

\section{derivatives of $\Omega_0$ }
By Eq($\ref{e1a}$), if $S$ is unchanged then $\bf{L}$ will
precession with a static velocity, $\Omega_{0}$, around $\bf{J}$.
However, since $\bf{S}_1$ and $\bf{S}_2$ precess with different
velocities, $\Omega_{1}$ and $\Omega_{2}$,  respectively. Thus
$\bf{S}$ varies both in magnitude and direction ($\bf{S}_1$,
$\bf{S}_2$ and $\bf{S}$ form a triangle), and in turn
$\lambda_{LJ}$ and  $\lambda_{LS}$ also vary with time
 (${\bf S}$, ${\bf L}$ and ${\bf J}$ form a
triangle, Fig.1).

 The motion of the vectors ${\bf S}_1$,
${\bf S}_2$ and ${\bf S}$ can be studied in the coordinate system
of the total angular momentum, in which the z-axis directs to
${\bf J}$, and the x- and y-axes are in the invariance plane.
${\bf S}$ can be represented by $S^{P}$ and $S^{V}$, the
components parallel and vertical to the z-axis, respectively:
\begin{equation}
\label{ex1}
 S=(S^{V}+S^{P})^{1/2} \ .
\end{equation}
$S^{P}$ and $S^{V}$ can be expressed  as
$$S^{P}=S_{1}\cos \lambda_{JS_1}+S_{2}\cos \lambda_{JS_2} \ ,$$
\begin{equation}
\label{ex2}  S^{V}= [(S^{V}_1)^2+(S^V_2)^{2}-2S^{V}_1S^{V}_2\cos
\eta_{S_1S_2}]^{1/2} \ ,
\end{equation}
$S_{1}^V=S_{1}\sin \lambda_{JS_1}$ and $S_{2}^V=S_{2}\sin
\lambda_{JS_2}$ represent components of $\bf{S}_{1}$ and
$\bf{S}_{2}$ that are vertical to the $\bf{J}$. $S^{V}_1$,
$S^{V}_2$ and $S^{V}$ form a triangle. $\eta_{S_1S_2}$, the
misalignment angle between $S^{V}_1$ and $S^{V}_2$ can be written
\begin{equation}
\label{ex5} \eta_{S_1S_2}=(\Omega_1-\Omega_2)t+\phi_{0}  \ .
\end{equation}
Note that $S$, $\lambda_{LS}$, $S_1^{\parallel}$
($S_1^{\parallel}=S_1\cos\eta_{SS_1}$), and
$\lambda^{\parallel}_{LS_1}$ in Eq($\ref{e1a}$) are function of
time, which leads to the change of the precession rate of orbit,
${\Omega}_{0}$.
\begin{equation}
\label{en1} \dot{\Omega}_{0} =\rho\dot{\Omega}+\dot{\rho}\Omega \
\,,
\end{equation}
where
\begin{equation}
\label{om01} \dot{\Omega}=\Omega_{2}\Omega_{12}X_3X_4-
\Omega_{12}X_1(\Omega_{01}X_2+\Omega_{12}X_3)+\Omega_{12}X_1X_5 \
\,,
\end{equation}
where $\Omega_{12}=\Omega_{1}-\Omega_{2}$,
$\Omega_{01}=\Omega_{1}-\Omega_{0}$,
 $$X_1=\frac{S^{\parallel}_{1}}{S}\sin \lambda_{LS_{1}} \,,
 X_2=\tan\eta_{ss1} \,,
 X_3=\frac{S_{1}^VS_{2}^V}{S^{2}}\frac{\sin\eta_{s1s2}}{\alpha\sin\lambda_{JS}}$$
$$ X_4=\frac{\cos^{2}\lambda_{LS}}{\sin \lambda_{LS}}   \ \,, \ \
X_5=\cot\lambda^{\parallel}_{LS_1}\dot{\lambda}^{\parallel}_{LS_1}
  \ \,, \ \
\dot{\rho}=-\frac{X_3X_4\Omega_{12}}{\sin^2\lambda_{JS}}$$ with
$\alpha=\sin \lambda_{JS}+\frac{\cos
^{2}\lambda_{LS}}{\sin\lambda_{LS}}$.

 Note that $\Omega_1$ and
$\Omega_2$ are unchanged, and $\lambda_{LS_{\alpha}}$ are
unchanged (since they decay much slower than the orbital
decay\cite{apo}. By Eq($\ref{en1}$), the second order derivative
of $\Omega_{0}$ are given,
\begin{equation}
\label{om2} \ddot{\Omega}_{0}
=\rho\ddot{\Omega}+2\dot{\rho}\dot{\Omega} +\ddot{\rho}{\Omega}\
\,,
\end{equation}
where
$$ \ddot{\Omega}
=\Omega_{2}\Omega_{12}(\dot{X}_{3}X_{4}+X_{3}\dot{X}_{4})-
\Omega_{12}\dot{X}_{1}(\Omega_{01}X_{2}+\Omega_{12}X_{3})-$$
\begin{equation}
\label{om22}\Omega_{12}X_{1}(\Omega_{01}\dot{X}_{2}+\Omega_{12}\dot{X}_{3})
+\Omega_{12}(X_{1}\dot{X}_{5}+\dot{X}_{1}X_5)
 \ \,,
\end{equation}
where
$$\dot{X}_1=-\Omega_{01}{X}_{1}\tan\eta_{SS_1}-(\Omega_{12}X_{2}X_{3})$$
$$\dot{X}_2=-\Omega_{01}\sec^2\eta_{SS_1}$$
$$\dot{X}_3=\dot{Y}\sigma+Y\dot{\sigma}$$
where $Y=\frac{S_{V1}S_{V2}\sin\eta_{S_1S_2}}{S^2}$ \,, \
$\sigma=\frac{1}{\alpha\sin\lambda_{JS}}$ \,, \ \
$$\dot{Y}=\Omega_{12}(Y\cot\eta_{S_1S_2}-Y^2\sigma)$$
$$\dot{\sigma}=-\frac{1}{(\alpha\sin\lambda_{JS})^2}
[\sin\lambda_{JS}\dot{X}_4$$
$$+(2\sin\lambda_{JS}+X_4)
\frac{\cos^2\lambda_{JS}}{\sin\lambda_{JS}}X_3\Omega_{12}]$$,
$$\ddot{\rho}=-\frac{\Omega_{12}}{\sin^2\lambda_{JS}}[\dot{X}_3X_4+\dot{X}_4X_3-
\frac{2\Omega_{12}X_3^2X_4^2}{\sin\lambda_{JS}}
 ]$$
$$\dot{X}_4=-\Omega_{12}X_3X_4(2+\frac{X_4}{\sin\lambda_{LS}})$$
$$\dot{X}_5=\ddot{\lambda}^{\parallel}_{LS_1}\cot\lambda^{\parallel}_{LS_1}-
(\dot{\lambda}^{\parallel}_{LS_1})^2\csc^2\lambda^{\parallel}_{LS_1}$$
where
$\lambda^{\parallel}_{LS_1}=\tan^{-1}(\tan\lambda_{JS_1}\cos\eta_{SS_1})$
\,, $\dot{\lambda}^{\parallel}_{LS_1}=\frac{\dot{z}}{1+z^2}$ \,,
$\ddot{\lambda}^{\parallel}_{LS_1}=\frac{\ddot{z}(1+z^2)-2z\dot{z}^2}{(1+z^2)^2}$
\ . In which $z=\tan\lambda_{JS_1}\cos\eta_{SS_1}$ \,,
$\dot{z}=-\Omega_{01}\tan\lambda_{JS_1}\sin\eta_{SS_1}$ \,,
$\ddot{z}=-\Omega^2_{01}\tan\lambda_{JS_1}\cos\eta_{SS_1}$ \,.

By Eq($\ref{om2}$), the third order derivative of $\Omega_{0}$ are
given,
\begin{equation}
\label{om3}  \frac{d^{3}{\Omega}_{0}}{d^{3}t} =
\Omega\frac{d^{3}{\rho}}{d^{3}t}+
\rho\frac{d^{3}{\Omega}}{d^{3}t}+3\ddot{\rho}\dot{\Omega}+3\dot{\rho}\ddot{\Omega}
\ \,,
\end{equation}
where
$$ \frac{d^{3}{\Omega}}{d^{3}t} =
\Omega_{2}\Omega_{12}(2\dot{X}_{3}\dot{X}_{4}+\ddot{X}_{3}X_{4}
+X_{3}\ddot{X}_{4})-$$
$$
2\Omega_{12}\dot{X}_{1}(\Omega_{01}\dot{X}_{2}+\Omega_{12}\dot{X}_{3})
-\Omega_{12}\ddot{X}_{1}(\Omega_{01}X_{2}+\Omega_{12}X_{3})-$$
\begin{equation}
\label{om33}
\Omega_{12}X_{1}(\Omega_{01}\ddot{X}_{2}+\Omega_{12}\ddot{X}_{3})
 +\Omega_{12}(X_{1}\ddot{X}_{5}+\ddot{X}_{1}X_5+2\dot{X}_{1}\dot{X}_5)\ \,,
\end{equation}
$$\ddot{X}_{1}=-\Omega_{01}\dot{X}_{1}\tan\eta_{SS_1}-\Omega_{01}^2X_{1}\sec^2\eta_{SS_1}$$
$$-\Omega_{12}\dot{X}_{1}X_3-\Omega_{12}\dot{X}_{3}X_1$$
$$\ddot{X}_{2}=2\Omega^2_{01}\tan\eta_{SS_1}\sec^2\eta_{SS_1}$$
$$\ddot{X}_{3}=\ddot{Y}\sigma+\ddot{\sigma}Y+2\dot{Y}\dot{\sigma}$$
$$\ddot{X}_{4}=-\Omega_{12}(2+\cot\lambda_{LS})(\dot{X}_{3}X_4+\dot{X}_{4}X_3)-$$
$$\Omega_{12}{X}_{3}X_4(\dot{X}_4\csc\lambda_{LS}
-\cos^2\lambda_{LS}\csc^3\lambda_{LS}\Omega_{12}{X}_{3}X_4)
$$
 where
$$\ddot{Y}=\Omega_{12}\dot{Y}\cot\eta_{S_1S_2}-\Omega^2_{12}{Y}\csc^2\eta_{S_1S_2}
-4\Omega_{12}Y\dot{Y}\sigma-2\Omega_{12}Y^2\dot{\sigma}
$$
$$\ddot{\sigma}=2\sigma\dot{\sigma}^2(\alpha\sin\lambda_{JS})^2
+\sigma^2[\sin\lambda_{JS}\ddot{X_4}+\dot{X_4}\dot{\lambda}_{JS}\cos\lambda_{JS}$$
$$+\Omega_{12}(2\dot{\lambda}_{JS}\cos\lambda_{JS}+\dot{X_4})\cos^2\lambda_{JS}\csc\lambda_{JS}X_3$$
$$+\Omega_{12}(2\sin\lambda_{JS}+{X_4})(\dot{\xi}X_3+\xi\dot{X}_3)]$$
$$\xi=\cos^2\lambda_{JS}\csc\lambda_{JS}$$
$$\dot{\xi}=-(2\cos\lambda_{JS}+\cos^3\lambda_{JS}\csc^2\lambda_{JS})\dot{\lambda}_{JS}$$
$$\dot{\lambda}_{JS}=\Omega_{12}X_3\cos\lambda_{JS}\csc\lambda_{JS}$$
$$\ddot{X}_5=2(\dot{\lambda}^{\parallel}_{LS_1})^2\cos\lambda^{\parallel}_{LS_1}
\csc^3\lambda^{\parallel}_{LS_1}-$$
$$3\dot{\lambda}^{\parallel}_{LS_1}\ddot{\lambda}^{\parallel}_{LS_1}
\csc^2\lambda^{\parallel}_{LS_1}+
\frac{d^3\lambda^{\parallel}_{LS_1}}{dt^3}\cot\lambda^{\parallel}_{LS_1}$$
where $\frac{d^3\lambda^{\parallel}_{LS_1}}{dt^3}=
\frac{(1+z^2)\frac{d^3z}{dt^3}-2z\dot{z}\ddot{z}-2\dot{z}^3
-4(1+z^2)z\dot{z}\ddot{\lambda}^{\parallel}_{LS_1}}{(1+z^2)^2}$ \
\ \ \,, \ \
$\frac{d^3z}{dt^3}=\Omega^3_{01}\tan\lambda_{JS_1}\sin\eta_{SS_1}$
\ \,.

\acknowledgements{I thank Prof. T.Y.Huang for helpful suggestions
and discussion.}

\clearpage

\begin{figure}[t]

\begin{center}
\includegraphics[87,87][700,700]{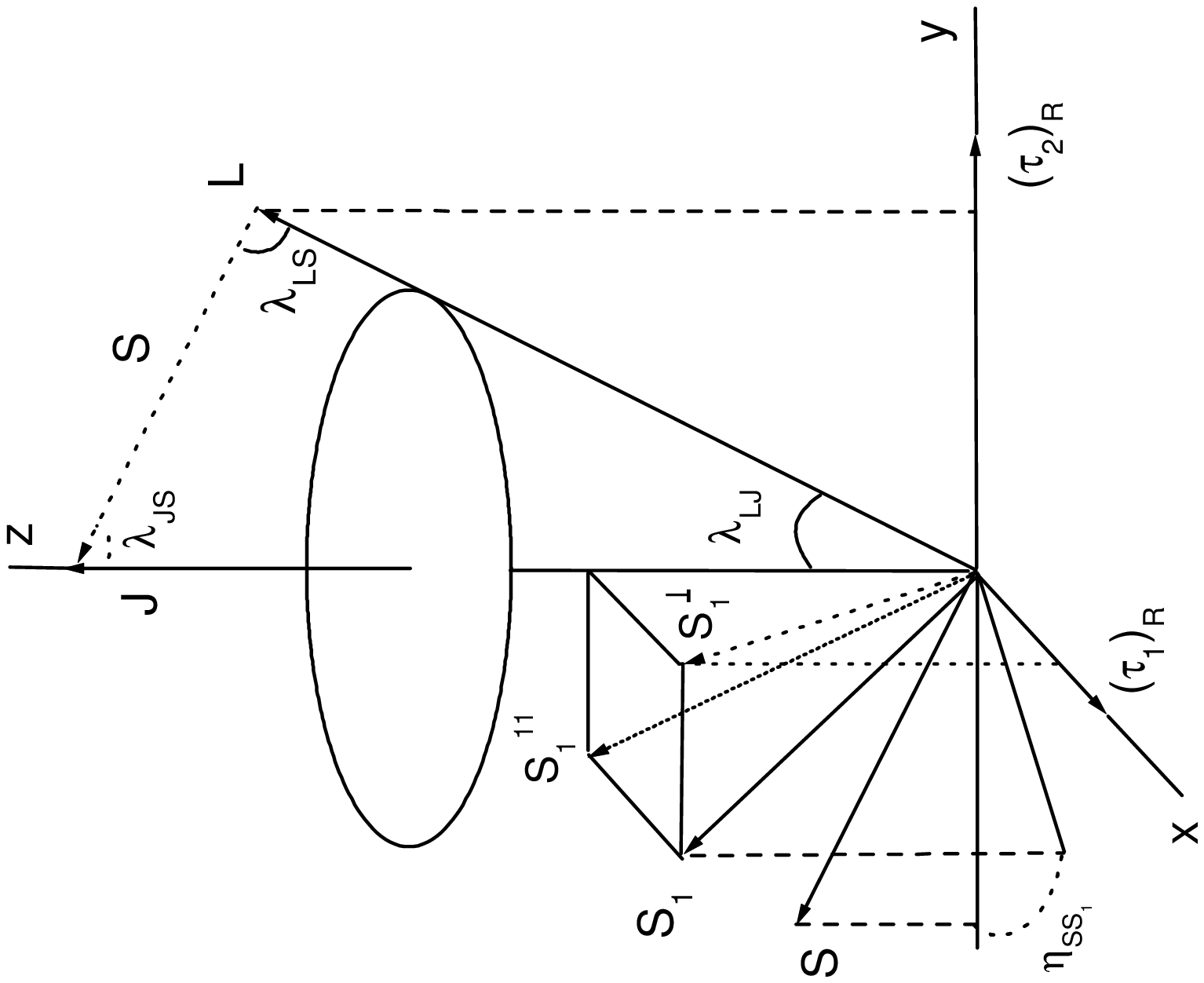}

\end{center}

\caption{Angles and orientation conventions relating
 vectors, ${\bf
S}$, ${\bf L}$ and ${\bf J}$ to the coordinate system. ${\bf S}$,
${\bf L}$ and ${\bf J}$ form a triangle, and determine a plane
LJS. At any instant ${\bf S}$ and ${\bf L}$ must be at the
opposite side of the fixed vector, ${\bf J}$ (invariable both in
direction and magnitude). $S^{\parallel}_1$ and $S^{\perp}_1$ are
components of ${\bf S}_1$ projected in the plane LJS and
perpendicular to it respectively. $(\tau_1)_R$ and  $(\tau_2)_R$
are torques corresponding to the precession and nutation of ${\bf
L}$ (or ${\bf S}$) in a binary system respectively. }
\end{figure}

\end{document}